Broadband Acoustic Hyperbolic Metamaterial

Chen Shen[1], Yangbo Xie[2], Ni Sui[1], Wenqi Wang[2], Steven A. Cummer[2] and Yun Jing[1,†]

[1]Department of Mechanical and Aerospace Engineering, North Carolina State University, Raleigh, North Carolina 27695, USA

[2]Department of Electrical and Computer Engineering, Duke University, Durham, North Carolina 27708, USA

†Corresponding author

yjing2@ncsu.edu

**Acoustic metamaterials (AMMs) are engineered materials, made from subwavelength structures, that exhibit useful or unusual constitutive properties. There has been intense research interest in AMMs since its first realization in 2000 by Liu *et al.*[1]. A number of functionalities and applications have been proposed and achieved using AMMs[2-10]. Hyperbolic metamaterials are one of the most important types of metamaterials due to their extreme anisotropy and numerous possible applications[11–13], including negative refraction, backward waves, spatial filtering, and subwavelength imaging. Although the importance of acoustic hyperbolic metamaterials (AHMMs) as a tool for achieving full control of acoustic waves is substantial, the realization of a broad-band and truly hyperbolic AMM has not been reported so far. Here, we demonstrate the design and experimental characterization of a broadband AHMM that operates between 1.0 kHz and 2.5 kHz.**

Acoustic hyperbolic metamaterials are AMMs that have extremely anisotropic densities. In two-dimensional (2D) scenarios, the density is positive in one direction and negative in the orthogonal direction. For materials with anisotropic densities, the general dispersion relation is given by $\frac{k_x^2}{\rho_x} + \frac{k_y^2}{\rho_y} = \frac{\omega^2}{B}$, where $k$ and $\omega$ are the wave number and angular frequency, respectively, $B$ is the bulk modulus of the medium. For media with positive, anisotropic densities, the equifrequency contour (EFC) is an ellipse, whereas it is a hyperbola for AHMMs. A broadband hyperlens has been demonstrated using brass fins[14]. The capability of subwavelength imaging, however, stems from extreme contrast of density and this hyperlens in fact does not bear a hyperbolic dispersion. It has been shown theoretically and experimentally that periodically perforated plates can exhibit hyperbolic-like dispersion for airborne sound[15,16]. Negative refraction and energy funneling associated with this structure were demonstrated experimentally within a narrow band around 40 kHz. This design yields a flat band profile in EFC and therefore could not demonstrate acoustic partial focusing, which is an important application of AHMMs. This is because under this band profile, the sound wave would travel in ballistic paths.

In this work, we show the realization of a broadband AHMM utilizing plate-(membrane-) type AMMs[17-22]. The proposed structure exhibits truly hyperbolic dispersion, as demonstrated by its ability of partial focusing and subwavelength imaging over a broadband frequency. The design of the AHMM is illustrated in Fig. 1. The rigid frames are made of aluminum and the thin plates are made of hard paper. There are 13 frames and each contains 14 plate unit cells. The boundaries of the plates are fixed securely on the aluminum frames to achieve the clamped boundary condition. No tension is applied on the plates. Two acrylic panels cover the top and bottom of the sample to ensure two-dimensional wave propagation.

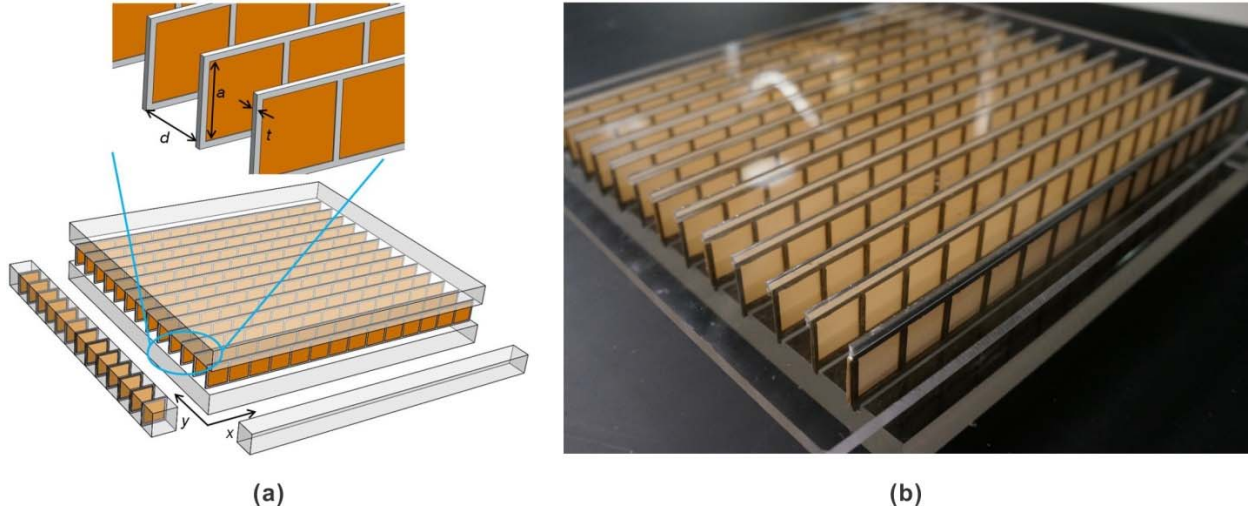

(a) (b)

**Figure 1 Snapshots of the AHMM. (a)** Physical structure of the AHMM. Each two frames have a separation distance of $d = 2$ cm and the thickness of the frame is $t = 0.16$ cm. The frames in the y-direction therefore have a periodicity of $D = d + t = 2.16$ cm. The width of the square plate is $a = 2$ cm. To study the effective acoustic properties, the 2D AHMM can be decoupled into two waveguides in each direction: one contains periodically arranged plates (y-direction) and the other does not (x-direction). **(b)** Photo of the fabricated AHMM sample. The size of the sample is 30.2 cm by 27 cm.

To theoretically characterize the proposed AHMM, we use one-dimensional (1D) analysis since wave propagation in x- and y-directions can be decoupled in this type of structure[18,20]. In other words, the effective acoustic property in x- and y-directions can be estimated by using 1D waveguide structures as shown in Fig.1. Because there are no plates arranged in the x-direction, the effective density can be considered as that of air and is frequency-independent ($\rho_x = 1.2$ kg/m$^3$). In the y-direction, a lumped model can be utilized to predict the frequency-dependent effective density[20], which is written as $\rho_y = \dfrac{Z_{am}}{j\omega} \cdot \dfrac{1}{DA}$, where $Z_{am}$ is the acoustic impedance of the plate, $A = a^2$ is the cross-section area of the waveguide. Since there is no closed form solution of $Z_{am}$ for a square plate, the acoustic impedance is calculated by the finite element method and is[20]

$$Z_{am} = \frac{Z_m}{A^2} = \frac{\iint \Delta p A}{j\omega\xi A^2},$$ where $Z_m$ is the mechanical impedance of the plate, $\Delta p$ is the pressure difference across the plate, and $\xi$ denotes the average transverse displacement of the plate. The plate has a Young's modulus 2.61 GPa, Poission's ratio 0.33[22], density 591 kg/m³ and thickness 0.3 mm. The Young's modulus is retrieved from the resonance frequency testing of a single clamped square plate (see Supplementary Material Section I). The resonance frequency of a clamped square plate is given by[23] $f_0 = \frac{1.61h}{A}\sqrt{\frac{E}{\rho(1-\nu^2)}}$. For the designed plate, the resonance frequency (or cutoff frequency) is about 2.68 kHz. Figure 2 shows the predicted effective density and the corresponding EFC as a function of frequency. $\rho_y$ is close to zero around the cutoff frequency and negative below the cutoff frequency[17,20,24]. Since $\rho_x$ is always positive, the dispersion curve theoretically is a hyperbola over a broadband frequency.

When the acoustic waves (red solid line) propagate from free space (black dotted curve) into the hyperbolic medium (purple solid curve) at a certain angle, the refractive angle is negative, since the group velocity $v_g$ must lie normal to the EFC. As a result, partial acoustic focusing can be achieved[13]. By combining the dispersion relation of free space and the hyperbolic medium, the refraction angle of an incident wave vector $k_i$ with vertical component $k_y$ can be calculated as:

$$\theta_t = \pm\tan^{-1}\frac{\rho_x}{\rho_y}\left(k_y \bigg/ \left(\rho_x\left(\frac{\omega^2}{B} - \frac{k_y^2}{\rho_y}\right)\right)^{\frac{1}{2}}\right). \quad (1)$$

The signs should be determined so that the refraction angle has the opposite sign with $k_y$. As seen from Fig. 2a, at frequencies sufficiently below the cutoff frequency, $\rho_y$ becomes deep negative. According to Eq. (1), the resulting refraction angle would approach zero, indicating that the focusing effect cannot be observed for an AHMM with a finite size. Since the fabricated AHMM is about 1-2 wavelengths at frequencies of interest, partial focusing effect can be best observed when the absolute value of $\rho_y$ is comparable to the background medium, which occurs at frequencies relatively close to the cutoff frequency.

By rearranging terms in the dispersion equation, one can see that $k_x^2 = \rho_x \left( \dfrac{\omega^2}{B} - \dfrac{k_y^2}{\rho_y} \right)$, which indicates that in the absence of losses, there does not exist a value for $k_y$ so that $k_x = 0$ since $\rho_y$ is negative and $\rho_x$ is positive. Consequently, all waves inside the AHMM are in propagating mode and no evanescent solutions are allowed. In other words, for an arbitrary incident angle, the evanescent wave reaching the surface of the AHMM can excite the propagating mode. Furthermore, the EFC becomes flat at frequencies sufficiently below the cutoff frequency, as the absolute value of the effective density in the y-direction is large (Fig. 2a). It can be predicted from Fig. 2b that at low frequencies, the refracted acoustic waves would experience almost ballistic travelling inside the AHMM without interaction. This implies subwavelength information can be retained on the opposite side of the AHMM and subwavelength imaging is possible.

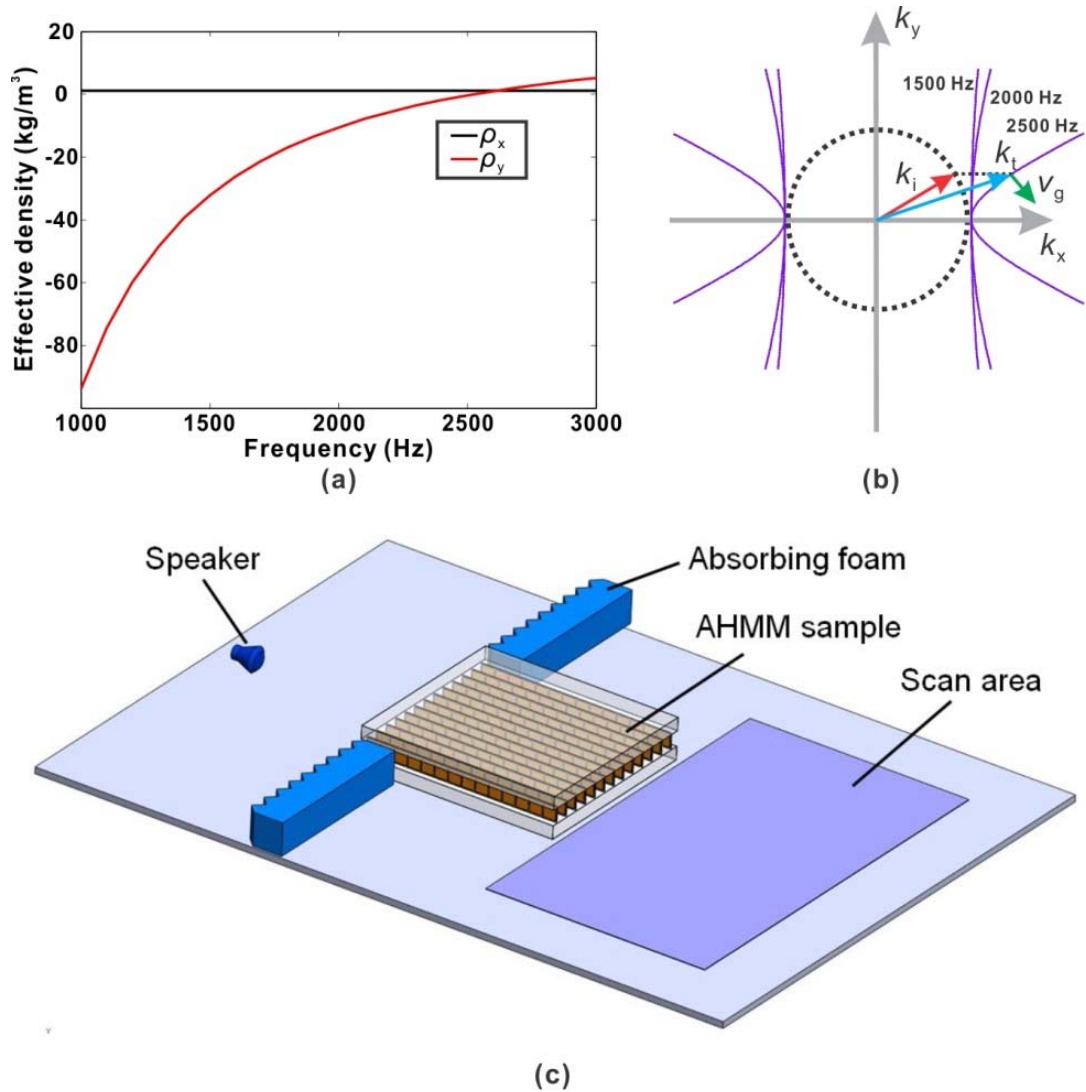

**Figure 2 Material properties of the AHMM and schematic of the experimental setup. (a)** Effective density along x- and y-directions. $\rho_y$ is retrieved from the lumped model and is negative below the cutoff frequency. **(b)** Calculated EFC at three selected frequencies which are all below the cutoff frequency. The dispersion curves are clearly hyperbolic and the EFCs become flat at low frequencies. **(c)** A loudspeaker mimicking a point source is placed 170 mm away from the front face of the sample. Sound absorptive materials are placed on the edges of the 2D waveguide

and two sides of the AHMM sample to minimize the reflection and sound field interference behind the AHMM, respectively.

Both numerical simulations and experiments are conducted to validate the proposed AHMM for partial focusing first. The setup for the experiment is depicted in Fig. 2c. Figure 3a shows the simulated and measured acoustic pressure fields at 2440 Hz for both cases where the AHMM is present and absent. The acoustic energy is focused on the back of the AHMM and diverges behind the AHMM, as shown by both simulations and experiments. The measurement agrees well with the simulation in terms of the pressure pattern. The pressure magnitude distributions on the exiting surface of the AHMM are also examined. Two types of simulations are performed. One uses the real structure of the AHMM and the other one uses effective medium with properties given by Fig. 2a. Although only partial focusing is achieved, the AHMM may still be favorable over an isotropic negative index metamaterial in terms of energy focusing as it is less sensitive to material loss[25]. When the frequency is above the cutoff, $\rho_y$ becomes positive and the AMM changes from a hyperbolic one to an elliptical one. The results at such a frequency can be found in Supplementary Material Section II. To further validate the effective medium model, which is the theoretical basis for designing the proposed AHMM, quantitative analysis is conducted for negative refraction in a long AHMM slab and the results can be found in Supplementary Material Section III.

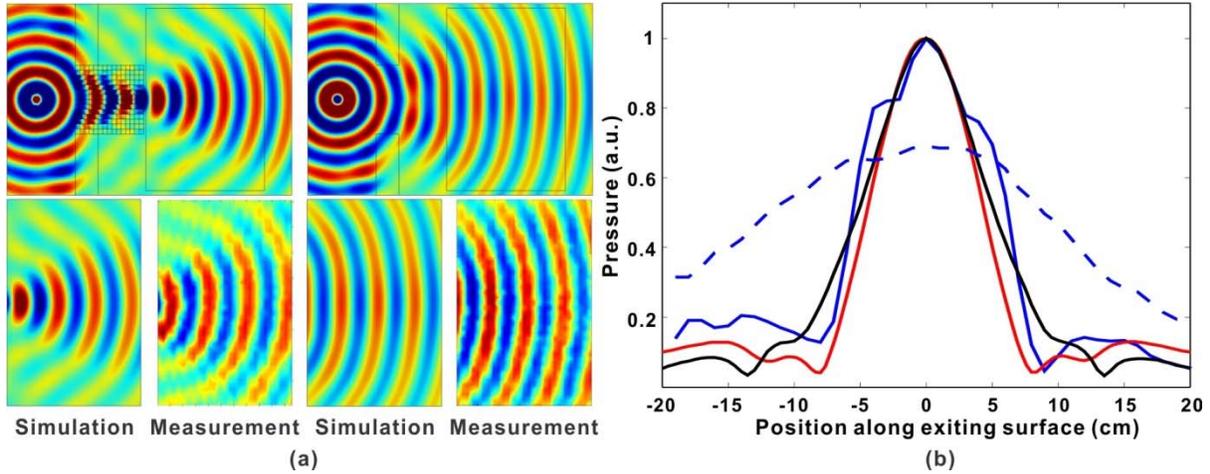

**Figure 3 Simulated and measured acoustic field showing partial focusing. (a)** Acoustic pressure field at 2440 Hz. Top two figures show the simulation results for the entire domain. Left one is with AHMM and right one is without AHMM. Bottom figures compare the simulation and measurement in the scanning area. **(b)** Normalized pressure magnitude distribution on the exiting surface of the AHMM. A focused profile can be clearly observed. Blue (solid): measurement with AHMM; blue (dashed): measurement without AHMM; red: simulation (real structure); black: simulation (effective medium).

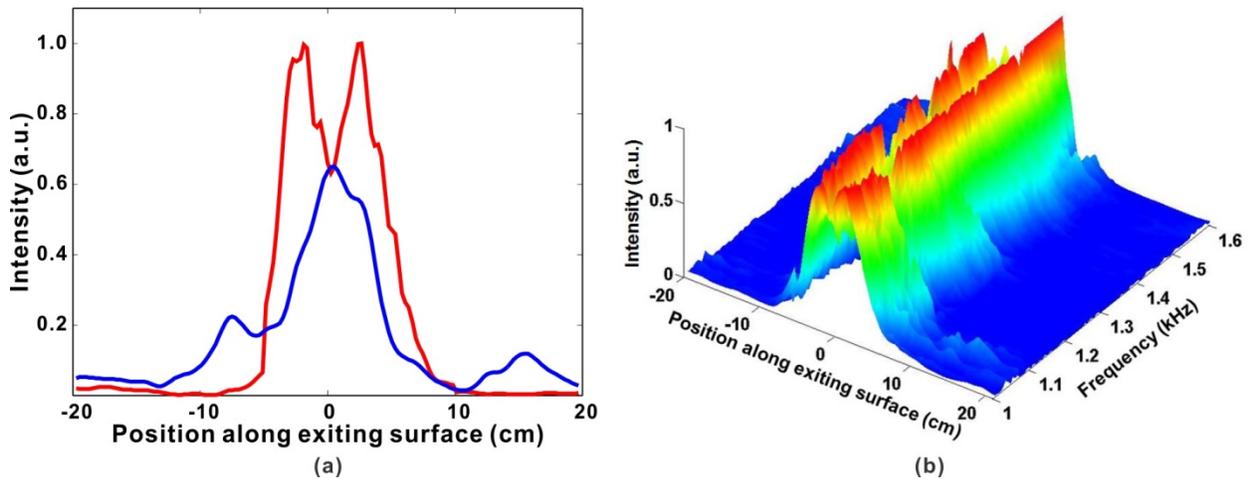

**Figure 4 Measured acoustic fields demonstrating subwavelength imaging. (a)** Imaging performance of the AHMM at 1.1 kHz. At this frequency, the resolved resolution is about 1/4.7 of the wavelength. The normalized acoustic intensity distribution along the exiting surface of the AHMM clearly shows two peaks, while the control case (without AHMM) shows a single peak. **(b)** The broadband performance of subwavelength imaging of the AHMM. Two peaks are resolved within a broad frequency band (1 kHz-1.6 kHz).

The AHMM is also capable of subwavelength imaging. As abovementioned, this is possible due to the flat EFCs, which occurs at frequencies sufficiently below the cutoff frequency. To demonstrate subwavelength imaging, a perforated panel with two square holes is placed in front of the AHMM, creating two in-phase sources separated by approximately 66 mm. The acoustic field is measured on the exiting surface of the AHMM at frequencies between 1 kHz and 1.6 kHz. The separation distance therefore corresponds to $\lambda/5.2$-$\lambda/3.2$. The thickness of the AHMM corresponds to $0.9\lambda$-$1.4\lambda$. Experimental results are shown in Fig. 4. At 1.1 kHz, two peaks are clearly resolved when the AHMM is present (Fig. 4a). When the AHMM is absent (control case), the waves radiated by the two sources merge and the resulting acoustic pressure field shows only one peak. The amplitudes of the peaks are also magnified when the AHMM is present compared to the control case. This is because most of the energy of the evanescent components, containing subwavelength information, is coupled into propagating modes and transferred to the image plane through the AHMM. Figure 4b demonstrates the broad-band performance of the AHMM. Two peaks can be observed for all frequencies within the frequency range tested.

To conclude, we have designed, fabricated, and tested a broadband AHMM based on plate-type AMMs. Partial focusing and subwavelength imaging are experimentally demonstrated within a broad frequency band, which verifies that such an AMM yields a truly hyperbolic dispersion. The proposed AHMM may find usage in angular filtering[26], medical imaging, non-destructive testing, etc. The proposed design can be readily extended to achieve three-dimensional AHMMs. We expect that the results of this paper will provide a new design methodology for the realization of AMMs requiring anisotropic, negative densities.

**Methods**

**Experimental method.** The measurement is conducted inside a 2D waveguide. The loudspeaker transmits pulsed signals at various center frequencies with a bandwidth of 1 kHz and the pressure fields are measured behind the sample. The field mapping measurements are performed using a 10 mm microphone mounted on a scanning stage. The scanning step size is 20 mm and at each position the acoustic pressures are averaged over five measurements. After scanning, the frequency-domain acoustic fields are obtained via the Fourier transform.

**Simulation method.** The commercial package COMSOL MULTIPHYSICS 5.1 is adopted for simulations. Acoustic-Shell interaction module is used to model the AHMM. All the simulations are performed in the frequency domain. The boundaries of the plates are set to be fixed and no tension is applied. The simulation setup is the same to that of the experiment. Absorptive materials are built on top and bottom of the AHMM to represent the sound absorbing foam. Perfectly matched layer (PML) is used to reduce the reflection from the boundaries. A point source is placed in front of the AHMM to generate the sound.

## Acknowledgements

This work was partially supported by the Multidisciplinary University Research Initiative grant from the Office of Naval Research (N00014-13-1-0631).

## Author contributions

C. S., Y. X, and W. W. conducted the experiments. C. S. performed the simulation and theoretical analysis. N. S. fabricated the sample. S. A. C. and Y. J. prepared the manuscript. All authors contributed to discussions.

## Additional information

Supplementary information is available in the online version of the paper. Reprints and permissions information is available online at www.nature.com/reprints. Correspondence and requests for materials should be addressed to Y. J or S. A. C.

## Competing financial interests

The authors declare no competing financial interests.

Supplementary Material

Broadband Acoustic Hyperbolic Metamaterial

Chen Shen[1], Yangbo Xie[2], Ni Sui[1], Wenqi Wang[2], Steven Cummer[2] and Yun Jing[1]

[1]Department of Mechanical and Aerospace Engineering, North Carolina State University, Raleigh, North Carolina 27695, USA

[2]Department of Electrical and Computer Engineering, Duke University, Durham, North Carolina 27708, USA

I. DETERMINATION OF THE YOUNG'S MODULUS OF THE PAPER PLATE

To determine the Young's modulus of the paper plates, a single square plate with width $a = 15.7$ mm is fabricated with boundaries clamped to an aluminum frame. The transmission loss (TL) of the sample is measured using a B&K impedance tube and is shown in Fig. S1. The TL dip observed at 4360 Hz indicates the resonance of the plate. From the equation $f_0 = \frac{1.61h}{A}\sqrt{\frac{E}{\rho(1-v^2)}}$, the Young's modulus can be estimated by:

$$E = \frac{f_0^2 A^2 \rho(1-v^2)}{2.59h^2} . \tag{1S}$$

The resulting Young's modulus $E$ is then 2.61 GPa.

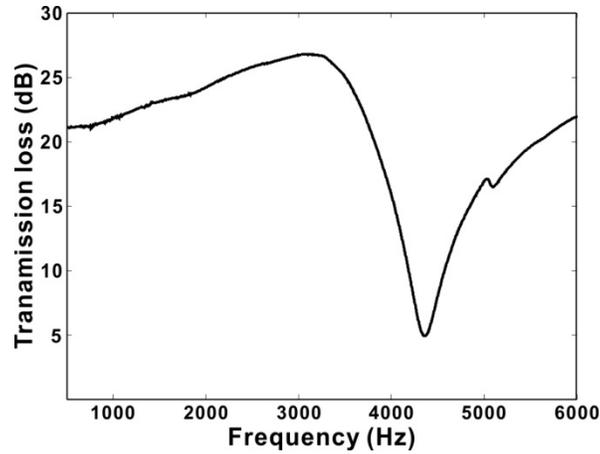

FIG. S1 Transmission loss of the plate sample. The resonance frequency is at 4360 Hz.

## II. SIMULATION AND MEASUREMENT RESULTS AT A FREQUENCY ABOVE THE CUTOFF FREQUENCY

At frequencies higher than the cutoff frequency of the plate, $\rho_y$ becomes positive and the refractive index changes from a negative value to a positive one. The resulting acoustic field will no longer show a focused profile. Results at 2.9 kHz are shown in Fig. S2 with and without the AHMM. Indeed, the energy focusing cannot be observed at this frequency.

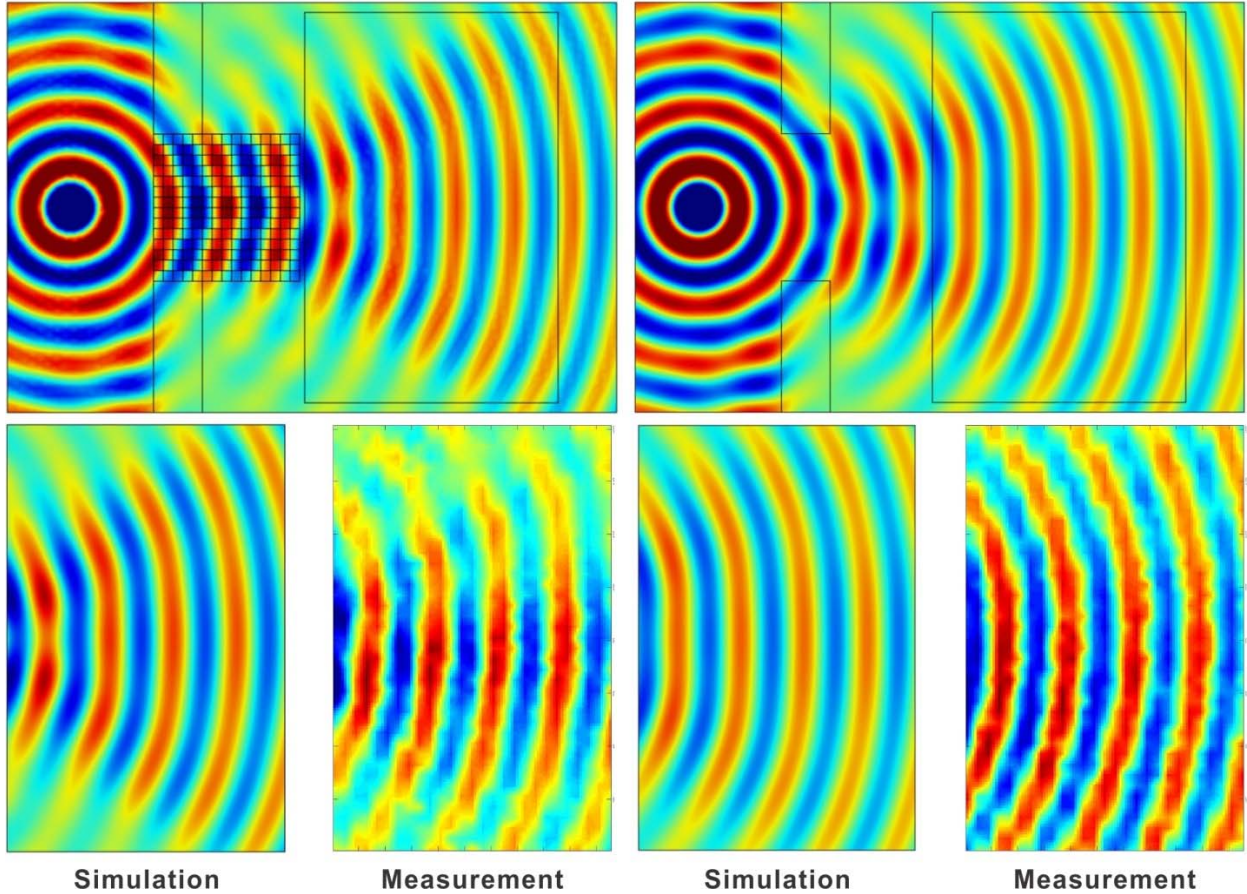

FIG. S2 Simulated and measured acoustic fields at 2900 Hz.

III. NEGATIVE REFRACTION

The effect of negative refraction can be demonstrated by a long AHMM slab placed in front of a Gaussian beam. For quantitative analysis, the shifted distance of the beam is estimated and compared between the real structure model and effective medium model.

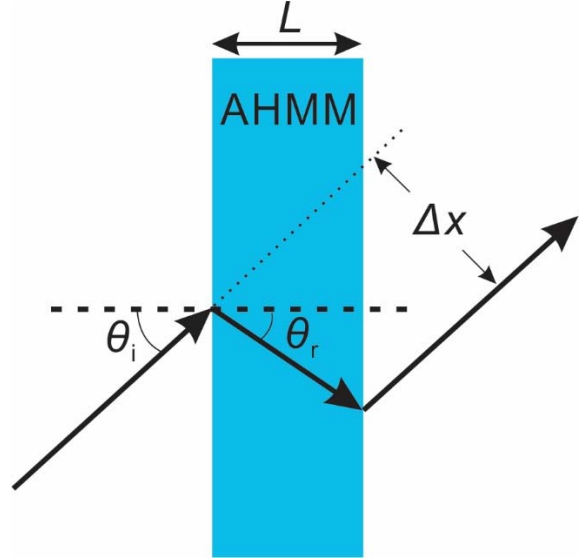

FIG. S3 The simulation setup for negative refraction. This behavior can be quantitatively evaluated by examining the shifted distance $\Delta x$ of the beam.

For an incident beam with an incident angle of $\theta_i$, the shifted distance can be calculated as:

$$\Delta x = \frac{L\cos(\theta_i - \theta_r)}{\cos\theta_r}, \qquad (2S)$$

where the refraction angle $\theta_r$ can be estimated by Eq. (5). Equation (2S) provides the analytically predicted shifted distance for the effective medium model. For comparison, the real structure of AHMM is simulated in COMSOL and the shifted distance is estimated as well.

The pressure field showing negative refraction using real structure is depicted in Fig. S4(a) at a frequency of 2450 Hz. The thickness of the slab is $L = 31.54$ cm (same with the fabricated sample). Figure S4(b) shows the theoretical and simulated shifted distance for an AHMM with a beam incident angle of $\theta_i = 45°$. The results agree well, both demonstrating a broadband negative refraction phenomenon below the cutoff frequency of the plates.

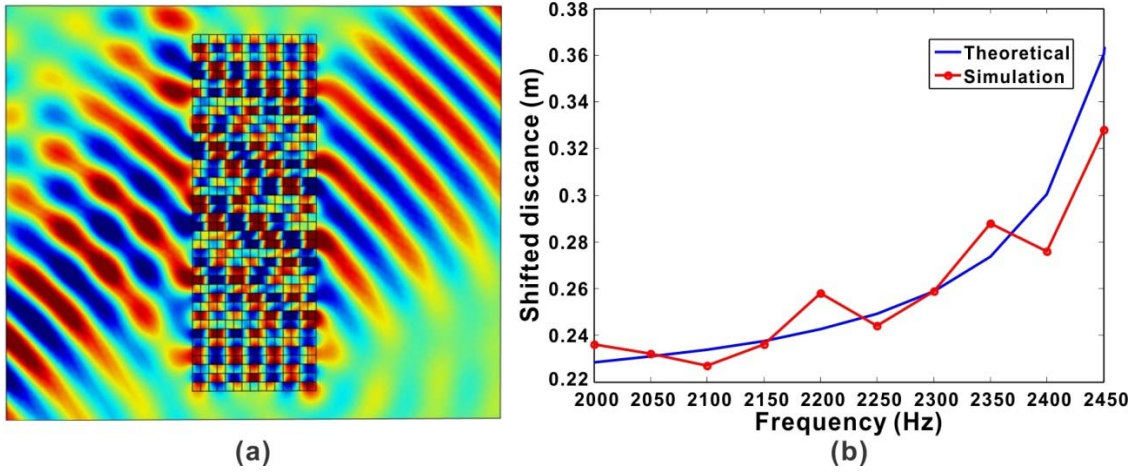

FIG. S4 (a) Numerical demonstration of negative refraction using COMSOL at 2450 Hz. (b) Theoretical and simulated shifted distances. The distance gradually increases as the frequency goes higher, as $\rho_y$ approaches zero and the corresponding refraction angle becomes greater.